\documentclass[12pt]{article}
\usepackage{epsfig}
\usepackage{hyperref}%

\date{}
\begin{document}

\title{Inhomogeneous Dark Radiation Dynamics on a de Sitter Brane}

\author{Rui Neves\footnote{E-mail: \tt rneves@ualg.pt}\hspace{0.2cm}
  and
Cenalo Vaz\footnote{E-mail: \tt cvaz@ualg.pt}\\
{\small \em \'Area Departamental de F\'{\i}sica/CENTRA, FCT,
Universidade do Algarve}\\
{\small \em Campus de Gambelas, 8000-117 Faro, Portugal}
}

\maketitle

\begin{abstract}
Assuming spherical symmetry we analyse the dynamics of an
inhomogeneous dark radiation vaccum on a Randall and Sundrum 3-brane
world. Under certain natural conditions we show that the effective
Einstein equations on the brane form a closed system. On a de Sitter
brane and for negative dark energy density we determine exact
dynamical and inhomogeneous solutions which depend on the brane
cosmological constant, on the dark radiation tidal charge and on its
initial configuration. We also identify the conditions leading to the
formation of a singularity or of regular bounces inside the dark
radiation vaccum.
\end{abstract}

\section{Introduction}  
                   
In the context of the intensive search for extra dimensions the
Randall and Sundrum (RS) brane world scenario is specially compelling for
its simplicity and depth \cite{RS}. In this scenario the observable
Universe is a 3-brane boundary of a non-compact $Z_2$ symmetric  
5-dimensional Anti-de Sitter (AdS) space. The matter fields are restricted to 
the brane but gravity exists in the whole AdS bulk and is bound to the
brane by the warp of the infinite fifth dimension.

In recent years numerous studies have been conducted within the RS
scenario [see \cite{RM1} for a recent review and notation]. 
From an effective 4-dimensional point of view \cite{SMS,SSM,RM2} 
the interactions existing between the brane and the bulk lead to a
modification of the Einstein equations by a set of two distinct terms,
namely a local high energy embedding term generated by the matter
energy-momentum tensor and a non-local term induced
by the bulk Weyl tensor. The resulting equations have a complex
non-linear dynamics. For example the exterior vaccum of a collapsing
distribution of matter on the brane is now filled with modes
originated by the bulk Weyl curvature and can no longer be a static
space \cite{BGM,GD}.

Previous investigations of the RS scenario have been focused on static 
or homogeneous dynamical solutions. In this proceedings we analyse
some aspects of the inhomogeneous dynamics of a RS brane world vaccum 
[see \cite{RC} for more details].    

\section{Vaccum Einstein Equations on the Brane}

In the 4-dimensional effective geometric approach to the RS brane
world scenario first introduced by Shiromizu, Maeda and Sasaki
\cite{SMS,SSM,RM2}, the induced Einstein vaccum field equations
on the brane are given by

\begin{equation}
{G_{\mu\nu}}=-\Lambda{g_{\mu\nu}}-{{\mathcal{E}}_{\mu\nu}},\label{4defe}
\end{equation}                 
where $\Lambda$ is the brane cosmological constant and 
the tensor ${{\mathcal{E}}_{\mu\nu}}$ is the limit on the
brane of the projected 5-dimensional Weyl tensor. On one
hand the Weyl symmetries ensure it is a symmetric and
traceless tensor. On the other the Bianchi identities constrain it to
satisfy the conservation equations

\begin{equation}
{\nabla_\mu}{{\mathcal{E}}^\mu_\nu}=0\label{ce}.
\end{equation}

The tensor ${{\mathcal{E}}_{\mu\nu}}$ 
can be written in the following general form \cite{RM2}

\begin{equation}
{{\mathcal{E}}_{\mu\nu}}=-{{\left({{\tilde{\kappa}}\over{\kappa}}\right)}^4}
\left[{\mathcal{U}}
\left({u_\mu}{u_\nu}+{1\over{3}}{h_{\mu\nu}}\right)+{{\mathcal{P}}_{\mu\nu}}+{{\mathcal{Q}}_\mu}{u_\nu}+
{{\mathcal{Q}}_\nu}{u_\mu}\right],
\end{equation}
where $u_\mu$ such that ${u^\mu}{u_\mu}=-1$ is the 4-velocity field
and ${h_{\mu\nu}}={g_{\mu\nu}}+{u_\mu}{u_\nu}$ is the tensor which projects
orthogonaly to
$u_\mu$. The forms $\mathcal{U}$, ${\mathcal{P}}_{\mu\nu}$ and 
${\mathcal{Q}}_\mu$ represent different characteristics of the effects induced
on the brane by the free gravitational field in the bulk. Thus,
$\mathcal{U}$ is interpreted as an energy density,
${\mathcal{P}}_{\mu\nu}$ as stress and
${\mathcal{Q}}_\mu$ as energy flux. 

Since the 5-dimensional metric is unknown, in general
${\mathcal{E}}_{\mu\nu}$ is not fully determined on the brane
\cite{SMS, SSM}. As 
a consequence the effective 4-dimensional theory is not closed and to
close it we need simplifying assumptions about the bulk degrees of
freedom. For example we may consider a static and spherically
symmetric brane vaccum with ${{\mathcal{Q}}_\mu}=0$, 
${{\mathcal{P}}_{\mu\nu}}\not=0$ and $\mathcal{U}\not=0$ to find the 
Reissner-Nordstr\"om black hole solution on the brane \cite{DMPR}.    
Let us now show that it is possible to take a non-static spherically
symmetric brane vaccum with ${{\mathcal{Q}}_\mu}=0$, $\mathcal{U}\not=0$, 
${{\mathcal{P}}_{\mu\nu}}\not=0$ and still close the system of
dynamical equations. 

Consider the general,
spherically symmetric metric in comoving coordinates \, $(t,r,\theta,\phi)$,

\begin{equation}
d{s^2}={g_{\mu\nu}}
d{x^\mu}d{x^\nu}=-{e^\sigma}d{t^2}+A^2d{r^2}+{R^2}d{\Omega^2},
\label{met}
\end{equation}
where $d{\Omega^2}=d{\theta^2}+{\sin^2}
\theta d{\phi^2}$, $\sigma=\sigma(t,r)$, $A=A(t,r)$, $R=R(t,r)$ and
$R$ is interpreted as the physical spacetime radius. If the stress is 
isotropic then the traceless ${\mathcal P}_{\mu\nu}$ will have the 
general form 

\begin{equation} 
{{\mathcal P}_{\mu\nu}}={\mathcal{P}}\left({r_\mu}{r_\nu}-{1\over{3}}{h_{\mu\nu}}\right),
\end{equation}
where ${\mathcal{P}}={\mathcal{P}}(t,r)$ and $r_\mu$ is the unit radial vector, given in the
above metric by ${r_\mu}=(0,A,0,0)$. Then

\begin{equation}
{{\mathcal{E}}_\mu^\nu}={{\left({{\tilde{\kappa}}\over{\kappa}}\right)}^4}\mbox{diag}\left(\rho,-
{p_r},-{p_T},-{p_T}\right),
\end{equation}
where the energy density and pressures are, respectively, 
$\rho={\mathcal{U}}$,
${p_r}=(1/3)\left({\mathcal{U}}+2{\mathcal{P}}\right)$ and ${p_T}=(1/3)\left({\mathcal{U}}-{\mathcal{P}}
\right)$. Substituting in the conservation Eq. (\ref{ce}) we obtain
the following expanded system \cite{TPS}

\[
2\frac{\dot A}{A}\left(\rho+{p_r}\right)=-2\dot{\rho}-
4{{\dot{R}}\over{R}}\left(\rho+{p_T}\right),
\]
\begin{equation}
\sigma'\left(\rho+{p_r}\right)=-{p_r}'+4{{R'}\over{R}}({p_T}-{p_r}),\label{cee}
\end{equation}
where the dot and the prime denote, respectively, derivatives
with respect to $t$ and $r$. A synchronous solution is obtained by
taking the equation of state $\rho+{p_r}=0$, giving 
${\mathcal{P}}=-2{\mathcal{U}}$ with $\mathcal{U}$ having 
the dark radiation form 

\begin{equation}
{\mathcal{U}}={{\left({{\kappa}\over{\tilde{\kappa}}}\right)}^4}
{Q\over{R^4}},\label{KfB}
\end{equation}
where the dark radiation tidal charge $Q$ is constant. Consequently, 
 we obtain

\begin{equation}
{G_{\mu\nu}}=-\Lambda{g_{\mu\nu}}+{Q\over{R^4}}\left({u_\mu}{u_\nu}-
2{r_\mu}{r_\nu}+{h_{\mu\nu}}\right),\label{dem}
\end{equation}
an exactly solvable closed system 
which depends on the two free parameters $\Lambda$ and $Q$. Indeed, its
solutions can be written in the LeMa\^{\i}tre-Tolman-Bondi form  

\begin{equation}
d{s^2}=-d{t^2}+{{{R'}^2}\over{1+f}}d{r^2}+{R^2}d{\Omega^2},
\end{equation}
where $R$ satisfies

\begin{equation}
\dot{R}^2={\Lambda\over{3}}{R^2}-{Q\over{R^2}}+f.
\label{deq}
\end{equation}
The function $f=f(r)>-1$ is naturally interpreted as the energy inside 
a shell labelled by $r$ in the dark radiation vaccum and is fixed
by its initial configuration. 

\section{Inhomogeneous de Sitter Solutions}

Let us assume from now on that $Q<0$ and $\Lambda>0$ [note that
according to the recent supernovae data
$\Lambda\sim{10^{-84}}{{\mbox{GeV}}^2}$ - see {\it{e.g.}}
\cite{SND}]. Then we have a de Sitter brane with
negative dark energy density and the gravitational
field is confined to the vicinity of the brane. 
The marginally bound models corresponding to $f=0$ are the static
solutions

\begin{equation}
d{s^2}=-\left(1+{Q\over{R^2}}-{\Lambda\over{3}}{R^2}\right)d{T^2}+
{{\left(1+{Q\over{R^2}}-{\Lambda\over{3}}{R^2}\right)}^{-1}}d{R^2}+
{R^2}d{\Omega^2}
\end{equation}
and were used for the static exterior of a collapsing sphere of
homogeneous dark radiation \cite{BGM,GD}. When $\Lambda=0$ this 
corresponds to the zero mass limit of the tidal
Reissner-Nordstr\"om black hole solution on the brane
\cite{DMPR}. 

It is for $f\not=0$ that the dark radiation vaccum really becomes
dynamical. The solutions are better organized by a redefinition of
$f$, the function $\beta=(3/\Lambda)[3{f^2}/(4\Lambda)+Q]$. As an
explicit example take $\beta>0$. Then $|f|>2\sqrt{-Q\Lambda/3}$ and  

\begin{equation}
\left|{R^2}+{{3f}\over{2\Lambda}}\right|=\sqrt{\beta}\cosh\left[\pm 2
\sqrt{{\Lambda\over{3}}}t+{\cosh^{-1}}\left(
{{\left|{r^2}+{{3f}\over{2\Lambda}}\right|}
\over{\sqrt{\beta}}}\right)\right],\label{sol1}
\end{equation}
where $+$ or $-$ refer respectively to expansion or collapse. Since
$R$ is a non-factorizable function of $t$ and $r$ these
are new non-static, cosmological and intrinsically inhomogeneous exact 
solutions for negative dark energy dynamics on a spherically symmetric
de Sitter brane. 

\section{Singularities and Rebounces}

The dark radiation dynamics defined by Eq. (\ref{deq})
can lead to the formation of shell focusing singularities at
$R=0$ and of regular rebounce points at some $R\not=0$. To analyse 
these issues consider

\begin{equation}
{R^2}{\dot{R}^2}=V(R,r)={{\Lambda}\over{3}}{R^4}+f{R^2}-Q.
\end{equation}    
If for all $R\geq 0$ we find $V>0$ then a shell focusing singularity
forms at $R={R_s}=0$. On the other hand if there exists
$R={R_*}\not=0$ where $V=0$ then a regular rebounce point
forms at $R={R_*}$. For the dark radiation vaccum it is clear that no
more than two regular rebounce epochs can be
found. Since $\Lambda>0$ there is always a phase of continuous
expansion to infinity with ever increasing speed. For $\Lambda>0$ and
$Q<0$ other phases are possible depending on $f(r)$. To ilustrate
consider $\beta>0 \Leftrightarrow |f|>2\sqrt{-Q\Lambda/3}$. If
$f>2\sqrt{-Q\Lambda/3}$ then $V$ is
positive for all $R\geq 0$. There are no rebounce points 
and the dark radiation shells may either expand
continuously or collapse to a shell focusing singularity at ${R_s}=0$.
 
\begin{figure}[]
\setlength{\unitlength}{1cm}
\center{\epsfig{file=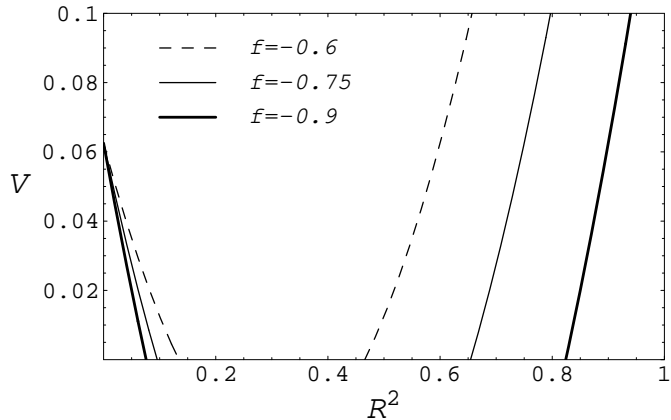,width=0.6\hsize}}
\caption{\small Plot of $V$ for $\beta>0$. Non-zero values of $f$ belong to the
interval $-1<f<-2\sqrt{-Q\Lambda/3}$ and correspond to
shells of constant $r$.}
\label{fig:Vppn}
\end{figure}

However for $-1<f<-2\sqrt{-Q\Lambda/3}$ 
(see Fig.~\ref{fig:Vppn})
there are two rebounce
epochs at $R=R_{*\pm}$ with ${R_{*\pm}^2}=-3f/(2\Lambda)\pm\sqrt{\beta}$.
Since $V(0,r)=-Q>0$ a singularity also forms at ${R_s}=0$. The
region between the two rebounce points is forbidden because there $V$ 
is negative. The phase space
of allowed dynamics is thus divided in two disconnected
regions separated by the forbidden interval ${R_{*-}}<R<{R_{*+}}$. For
$0\leq R\leq{R_{*-}}$ the
dark radiation shells may expand to a maximum radius $R={R_{*-}}$, 
rebounce and then fall to the singularity.
If $R\geq{R_{*+}}$ then there
is a collapsing phase to the minimum radius $R={R_{*+}}$ 
followed by reversal and subsequent accelerated continuous expansion. 
The singularity at ${R_s}=0$ does not form and so the solutions are globally
regular.

\section{Conclusions}

In this work we have reported some new results on the gravitational
dynamics of inhomogeneous dark radiation on a RS brane. Taking an 
effective 4-dimensional approach we have shown that under certain
simplifying but natural assumptions the Einstein field equations on the
brane form a closed, solvable system. We have presented exact dynamical and
inhomogeneous solutions for $\Lambda>0$ and $Q<0$ showing they
further depend on the energy function $f(r)$. We have also described the 
conditions under which a singularity or a regular bounce develop
inside the dark radiation vaccum. Left for future research are for
instance the analysis of the confinement of gravity near the brane and
the introduction of an inhomogeneous collapsing distribution of
matter.
\vspace{0.5cm}

\centerline{\bf Acknowledgements}
\vspace{0.25cm}

We thank the Funda\c {c}\~ao para a Ci\^encia e a Tecnologia (FCT) for
financial support under the contracts SFRH/BPD/7182/2001 and 
POCTI\-/32694/FIS/2000. We also would like to thank Louis Witten and
T.P. Singh for kind and helpful comments.

\end{document}